\documentclass[superscriptaddress,twocolumn,aps,prl,longbibliography]{revtex4-2}
\usepackage{amsmath,amssymb,graphicx,empheq,color}
\usepackage{natbib,dsfont,physics} %Added by Wouter
\usepackage[dvipsnames]{xcolor}
\usepackage{scalerel}

\def \a{a}
\def \ad{a^{\dagger}}
\def \Hami{\mathcal{H}}
\def \Green{\mathcal{G}}
\def \V{\mathcal{V}}
\def \dm{\rho}

\newcommand{\sddots}{\scaleobj{0.7}{\ddots}}
\newcommand{\svdots}{\scaleobj{0.7}{\vdots}}
\newcommand{\shdots}{\scaleobj{0.7}{\hdots}}

\begin{document}
\title{Giant Enhancement of Unconventional Photon Blockade in a Dimer Chain}

\author{You~Wang}
\thanks{These authors contributed equally.}
\affiliation{Division of Physics and Applied Physics, School of Physical and Mathematical Sciences, Nanyang Technological University, Singapore 637371, Singapore}

\author{W.~Verstraelen}
\thanks{These authors contributed equally.}
\affiliation{Division of Physics and Applied Physics, School of Physical and Mathematical Sciences, Nanyang Technological University, Singapore 637371, Singapore}

\author{Baile Zhang}
\email{blzhang@ntu.edu.sg}
\affiliation{Division of Physics and Applied Physics, School of Physical and Mathematical Sciences, Nanyang Technological University, Singapore 637371, Singapore}
\affiliation{Centre for Disruptive Photonic Technologies, School of Physical and Mathematical Sciences, Nanyang Technological University, Singapore 637371, Singapore}

\author{Timothy C.~H.~Liew}
\email{timothyliew@ntu.edu.sg}
\affiliation{Division of Physics and Applied Physics, School of Physical and Mathematical Sciences, Nanyang Technological University, Singapore 637371, Singapore}
\affiliation{MajuLab, International Joint Research Unit UMI 3654, CNRS, Universit\'e C\^ote d'Azur, Sorbonne Universit\'e, National University of Singapore, Nanyang Technological University, Singapore}

\author{Y.~D.~Chong}
\email{yidong@ntu.edu.sg}
\affiliation{Division of Physics and Applied Physics, School of Physical and Mathematical Sciences, Nanyang Technological University, Singapore 637371, Singapore}
\affiliation{Centre for Disruptive Photonic Technologies, School of Physical and Mathematical Sciences, Nanyang Technological University, Singapore 637371, Singapore}

\date{\today}

\begin{abstract}
  Unconventional photon blockade refers to the suppression of multi-photon states in weakly nonlinear optical resonators via the destructive interference of different excitation pathways.  It has been studied in a pair of coupled nonlinear resonators and other few-mode systems.  Here, we show that unconventional photon blockade can be greatly enhanced in a chain of coupled resonators.  Specifically, the strength of the nonlinearity in each resonator needed to achieve unconventional photon blockade is suppressed exponentially with lattice size.  The analytic derivation, based on a weak drive approximation, is validated by wavefunction Monte Carlo simulations.  These findings show that customized lattices of coupled resonators can be powerful tools for controlling multi-photon quantum states.
\end{abstract}

\maketitle

Photon blockade---the use of optical nonlinearity to suppress multi-photon quantum states---is a mechanism for generating nonclassical light through classical optical illumination \cite{Tian92, Leo94, Imamo97, Lodahl15}, with applications in quantum computing, quantum simulation, and other emerging quantum technologies \cite{Knill01, Birnbaum05, Hartmann06, Greentree06, Kok07, Angelakis07, Umucal12,Noh16}.  The conventional photon blockade effect requires strong optical nonlinearities, as it relies on interactions between resonant single-photon states and off-resonant multi-photon states, so the interaction strength has to be much larger than the cavity decay rate.  This regime can be achieved in cavity QED systems \cite{Birnbaum05, Dayan08, Hamsen17, Michler00, Faraon08, Claudon10, He13, Madsen14, Gschrey15, Somaschi16, Dory17, Jia18}, superconducting circuits \cite{Lang11, Wang16}, optomechanical resonators \cite{Rabl11, Wang16, Xu16, Lemonde16}, and other systems \cite{Majumdar12, Majumdar13}.  Weak nonlinearities, however, are much easier to realize, such as in resonators made of common nonlinear optical materials.  Remarkably, it is possible to efficiently suppress multi-photon states even in the weakly nonlinear regime, through the phenomenon of unconventional photon blockade (UPB).  Liew and Savona showed some years ago that in a system of two coupled nonlinear resonators, careful parameter tuning can enable destructive interference between different excitation pathways for the formation of two-photon states in the signal resonator, even when the photon interaction strength is smaller than the cavity decay rate \cite{Liew10}.  Subsequently, the conditions for UPB to occur have been extensively studied \cite{Ferretti10, Bamba11, Bamba11_2, Flayac13, Xu14, Lemonde14}, and the phenomenon has been realized in experiments \cite{Snijders18,Vaneph18}.  Other ways of realizing UPB using different setups and different quantum interference schemes have also been proposed \cite{Gerace14, Flayac15, Shen15, Flayac17, Gangcheng17, Sarma17, Ghosh18, Shen18, Sarma18, Ghosh19, Carmichael91, Bamba11, Radulaski17, Kamide17, Zubizarreta20}, and similar ideas have been explored for other forms of multi-photon state control in weakly nonlinear systems, such as for creating entangled photon sources \cite{Liew12, Liew13}.

In the context of classical optics and photonics, synthetic lattices such as photonic crystals \cite{John87, Yablonovitch87, Joannopoulos08} and photonic metamaterials \cite{Pendry00, Shamonina07, Valentine08, Cubukcu03} have proven to be versatile platforms for wave manipulation.  By offering a richer set of customizable degrees of freedom, such as lattice symmetries, they have the potential to outperform devices composed of individual or a few coupled optical cavities, or even access qualitatively different behaviors.  For instance, photonic lattices can exhibit bound states in the continuum, whose decay rates vanish due to destructive interference of numerous decay pathways \cite{Hsu16}.

In this Letter, we show that a lattice of coupled resonators can achieve UPB at much lower levels of optical nonlinearity than in previously-studied two-resonator setups.  We consider resonators arranged in a dimer chain, or Su-Schrieffer-Heeger (SSH) lattice \cite{SSH}, a one-dimensional model lattice whose single-particle properties have been extensively studied.  By extending the analysis to two-photon states and exploiting the lattice's various features, including its chiral symmetry, we derive analytic expressions for the one- and two-photon quantum amplitudes to leading order in the inter-cell coupling.  Remarkably, we find that the nonlinear Kerr coefficient necessary for UPB in a given signal resonator is suppressed exponentially in the total number of sites.  The theory predicts the necessary resonator frequency detunings and cavity decay rates, which form a striking pattern of complex roots in a 2D parameter space.  For the limiting case of just two resonators, our formulas reproduce previously-reported results \cite{Bamba11}. We have also performed wavefunction Monte Carlo (WFMC) simulations of the multi-photon system, which give good agreement with the analytic results, and help quantify the limits on photon antibunching imposed by pure dephasing.  Recently, two-mode UPB has been demonstrated with quantum dot cavities \cite{Snijders18} and superconducting circuits \cite{Vaneph18}, and our findings may help in designing lattice-based single-photon sources on weakly-nonlinear platforms such as silicon photonics \cite{Flayac15,Ferretti2013}.

\begin{figure}
  \includegraphics[width=0.95\linewidth]{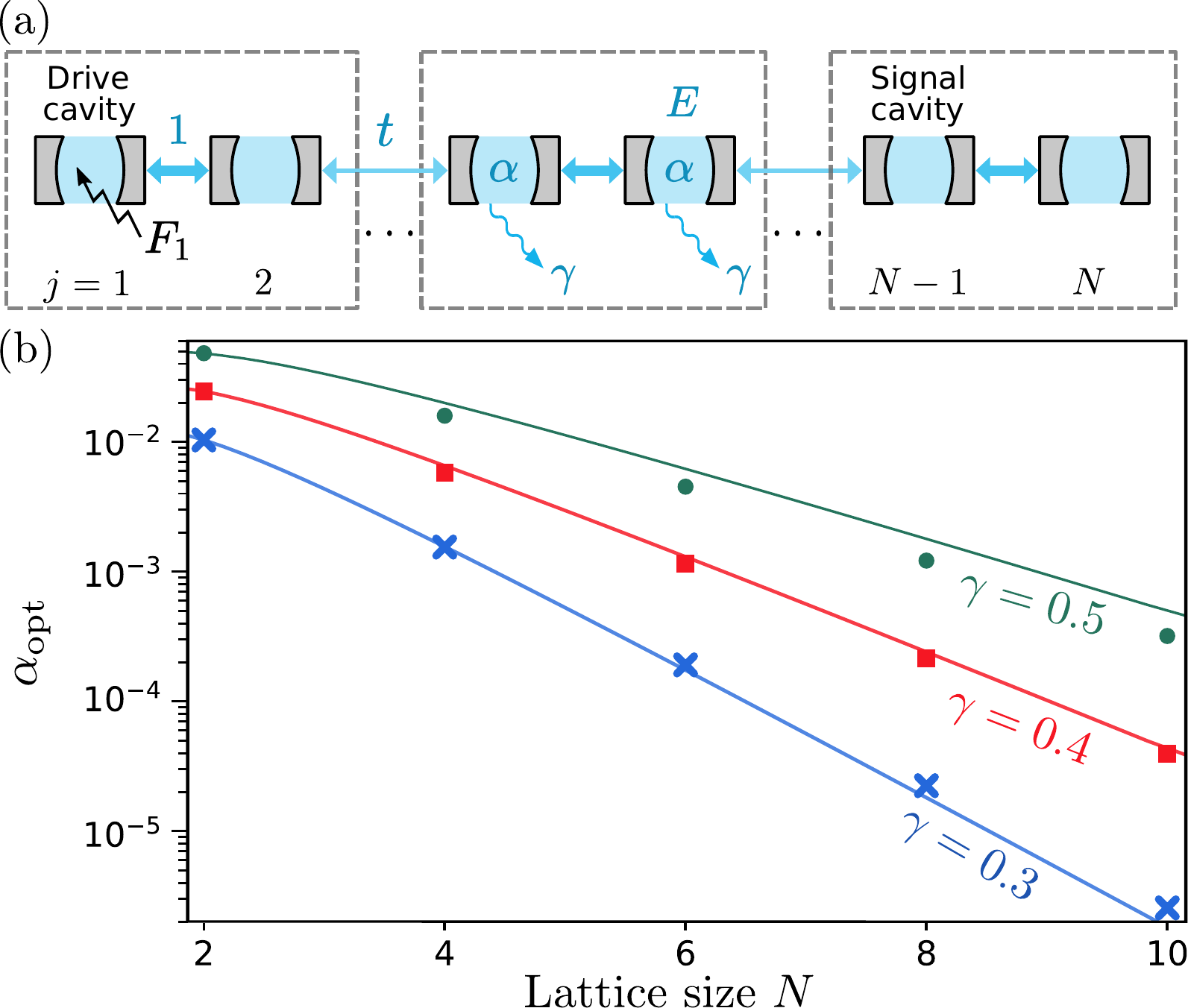}
  \caption{(a) Schematic of a dimer chain lattice of optical cavities. The cavity on site $j = 1$ is coherently driven, and site $N-1$ is the signal resonator.  Each cavity has single-photon energy $E$, Kerr coefficient $\alpha$, and decay rate $\gamma$.  Dashed boxes indicate the unit cells; the intra- and inter-cell couplings are $1$ and $t$ respectively.  (b) Optimum Kerr coefficient for different lattice lengths with $t=0.1$ and $\gamma=0.3, \, 0.4, \, 0.5$.  The solid curves correspond to the analytic approximation \eqref{eq:alpha_scaling}, while the discrete data points are calculated numerically from the weak drive equations.}
  \label{fig:lattice}
\end{figure}

Consider coupled optical resonators with identical physical properties and weak Kerr-type nonlinearity, arranged in a dimer chain (or SSH lattice \cite{SSH}) as shown in Fig.~\ref{fig:lattice}(a).  The number of sites, $N$, is even.  In the absence of driving and dissipation, the Hamiltonian is $\Hami_0 = \Hami_c+\Hami_p+\Hami_{nl}$, where
\begin{align}
  \begin{aligned}
    \Hami_c &= \left(\sum_{j=1}^{N/2}\ad_{2j-1}\a_{2j}
    + t\sum_{j=1}^{N/2-1}\ad_{2j}\a_{2j+1}\right) + \textrm{h.c.} \\
    \Hami_p &= E \sum_{j=1}^{N}\ad_j\a_j, \qquad
    \Hami_{nl} = \alpha\sum_{j=1}^N \ad_j\ad_j\a_j\a_j.
  \end{aligned}\label{eq:H}
\end{align}
Here, $\a_j^{(\dagger)}$ is the photon annihilation (creation) operator on site $j$, $t$ is the intercell hopping (with the intra-cell hopping normalized to unity), $E$ is the on-site single-photon energy, $\alpha \in \mathbb{R}$ is the Kerr coefficient, and ``h.c.''~stands for the Hermitian conjugate. Here we choose positive couplings; negative couplings can be equivalently handled by redefining the operators on even sites. Note that although the SSH model is well-known for the existence of ``topologically protected'' single-particle eigenstates at boundaries and domain walls, this behavior is not used in the present work; we exploit its other features, such as the chiral symmetry of the lattice.  We consider a chain with no domain walls, and (since $N$ is even) no single-particle topological eigenstates.

Let the sites be coherently driven by the Hamiltonian
\begin{equation}
  \Hami_d = \sum_{j=1}^N F_j\ad_j + \mathrm{h.c.},
  \label{eq:H_d}
\end{equation}
where $F_j$ is the excitation coefficient on site $j$.  Fig.~\ref{fig:lattice}(a) depicts the case where only the first site is driven, $F_j = F_1 \delta_{1j}$.  The evolution of the density matrix $\dm$ is given by the Lindblad master equation \cite{BreuerBookOpen,*Carmichael_BOOK_2}
\begin{equation}
  i\hbar\frac{d\dm}{dt} = \left[\Hami_{\mathrm{tot}},\dm\right]
  + \frac{i\gamma}{2}\sum_{j=1}^N
  \left(2\a_j\dm \, \ad_j - \ad_j\a_j\dm - \dm\,\ad_j\a_j\right),
  \label{eq:Lindblad}
\end{equation}
where $\Hami_{\mathrm{tot}} = \Hami_0 + \Hami_d$.  The terms in parentheses represent the interaction of the system with the environment in the form of on-site losses, giving rise to (i) a deterministic decay and (ii) stochastic quantum jumps stemming from the fluctuation-dissipation theorem.  Both of these effects will be accounted for when we later solve Eq.~\eqref{eq:Lindblad} using stochastic WFMC simulations \cite{Dum92, Molmer93, Carmichael_1993, Barchielli_1991}. For now, however, we pursue an approximate analytic solution by neglecting the fluctuations, and absorbing the deterministic decay terms into the Hamiltonian.  Let us define
\begin{align}
  z = E-\frac{i\gamma}{2}, \quad \Hami_p'=z\sum_{j=1}^N \ad_j\a_j,
\end{align}
and consider the semiclassical regime where the time evolution is well described by the Schr\"odinger equation with the non-Hermitian Hamiltonian $\Hami = \Hami_c + \Hami_p' + \Hami_{nl}$.  Flayac and Savona have argued that this is valid in the ``weak drive'' limit $F_j\rightarrow0$, since stochastic jumps are rare when photon occupation numbers are low \cite{Flayac17}.

The steady state solution has the form
\begin{equation}
  \ket{\psi}=\sum_{k=0}^\infty |\psi^{(k)}\rangle,
\end{equation}
where $|\psi^{(k)}\rangle$ is the projection of the full wavefunction into the $k$-photon subspace.  In the weak drive limit, the amplitude for the higher photon number states is negligible.  Truncating at $k=2$, we obtain \cite{SM}
\begin{align}
  |\psi^{(1)}\rangle &= - \Hami^{-1}\Hami_{+} |\psi^{(0)}\rangle, \label{eq:steady1} \\
  |\psi^{(2)}\rangle &= - \Hami^{-1}\Hami_{+} |\psi^{(1)}\rangle, \label{eq:steady2}
\end{align}
where $\Hami_+=\sum_jF_j\, \ad_j$.  For $k = 1$, we adopt the eigenstate basis of $\Hami_c^{(1)}$ (the projection of $\Hami_c$ to the $1$-photon subspace), defined by
\begin{equation}
  \Hami_c^{(1)}\ket{\varphi_n}=\epsilon_n\ket{\varphi_n}.\label{eq:spectrum}
\end{equation}
Hence, the solution to Eq.~\eqref{eq:steady1} can be written as
\begin{equation}
  |\psi^{(1)}\rangle = \sum_n \frac{f_n\ket{\varphi_n}}{z+\epsilon_n}, \quad
  f_n = \sum_{j=1}^N F_j \braket{\varphi_n}{j}, \label{eq:psi1}
\end{equation}
where $|j\rangle \equiv \ad_j|\psi^{(0)}\rangle$.  Details of the derivation are given in the Supplemental Materials \cite{SM}.  Next, for $k = 2$, we define a basis formed by tensor products of the single-particle eigenstates, $\ket{\varphi_{mn}} \equiv \ket{\varphi_m}\otimes\ket{\varphi_n}$, and seek perturbative solutions to Eq.~\eqref{eq:steady2} of the form
\begin{equation}
  |\psi^{(2)}\rangle \approx |\psi_0^{(2)}\rangle + \alpha |\psi^{(2)}_1\rangle,
  \label{eq:psi2_perturbation}
\end{equation}
where the solution in the absence of interactions is
\begin{equation}
  | \psi_0^{(2)}\rangle = \frac{1}{\sqrt{2}}\sum_{mn}\frac{f_mf_n}{(z+\epsilon_m)(z+\epsilon_n)}\ket{\varphi_{mn}},
  \label{eq:psi2_0}
\end{equation}
and the perturbative correction can be shown to be \cite{SM}
\begin{align}
  |\psi_1^{(2)}\rangle = \!\!\sum_{imnpq}\!\! \ket{\varphi_{mn}} \frac{-\sqrt{2}f_{p}f_{q} \braket{\varphi_{mn}}{i,i} \braket{i,i}{\varphi_{pq}} }{(z+\epsilon_{p})(z+\epsilon_{q})(2z+\epsilon_m+\epsilon_n)}, \label{eq:psi2_1}
\end{align}
where $\ket{i,i} \equiv \ket{i}\otimes\ket{i}$.

We now let only the first site be driven, so that $f_n = F_1 \braket{\varphi_n}{1}$.  The $N = 2$ lattice, corresponding to a single dimer, has previously been shown \cite{Liew10} to exhibit UPB in site 1, which serves as both the drive and signal resonator.
%% UPB for $N=2$ lattice is related to a zero-energy two-photon eigenstate, which corresponds to two single-photon eigenstates of opposite eigenenergy~\cite{SM}.
For larger $N$, we will demonstrate enhanced UPB on a designated signal resonator on site $N-1$ of the chain (i.e., one site away from the end of the chain, opposite to the drive cavity), as shown in Fig.~\ref{fig:lattice}(a).  UPB shall be achieved if the equal time second order photon correlation in the signal resonator,
\begin{equation}
  g^{(2)}_{s}(0) = 2 \frac{\left|\braket{N-1,N-1}{\psi^{(2)}}\right|^2}{\left|\braket{N-1}{\psi^{(1)}}\right|^4},
  \label{eq:g2_def}
\end{equation}
vanishes.  Note that plugging only Eq.~\eqref{eq:psi2_0} into Eq.~\eqref{eq:g2_def} gives $g^{(2)}_{s}(0)=1$ (i.e., in the linear regime the emission is always coherent).

If the intercell coupling is weak ($t \ll 1$), we can estimate the two-photon state by applying Laurent series expansions to Eqs.~\eqref{eq:psi2_0} and \eqref{eq:psi2_1} in the domain $t < |z| < 1$.  The derivation, given in the Supplemental Materials \cite{SM}, utilizes the chiral symmetry of $\Hami_c^{(1)}$, which ensures that the single-photon spectrum is symmetric around $E = 0$.  The result is
\begin{align}
  \braket{N-1,N-1}{\psi_0^{(2)}}&\approx \frac{F_1^2}{\sqrt{2}}\, t^{N-2} \,z^2,
  \label{eq:Laurent_leading_term} \\
  \braket{N-1,N-1}{\psi^{(2)}_1} &\approx
  \frac{F_1^2}{\sqrt{2}}
  \frac{(-1)^{\frac{N}{2}+1}(N-3){!}{!}t^{N-2}}{(N-2){!}{!}(2z)^{N-1}}. \nonumber
\end{align}
Referring to Eq.~\eqref{eq:psi2_perturbation}, UPB is achieved when 
\begin{equation}
  \alpha \approx \frac{(-1)^{\frac{N}{2}}}{4} \, \frac{(N-2){!}{!}}{(N-3)!!} \,
  (2z)^{N+1}. \label{eq:alpha_z_final}
\end{equation}
Notably, this is independent of the drive amplitude $F_1$.

\begin{figure}
  \includegraphics[width=0.95\linewidth]{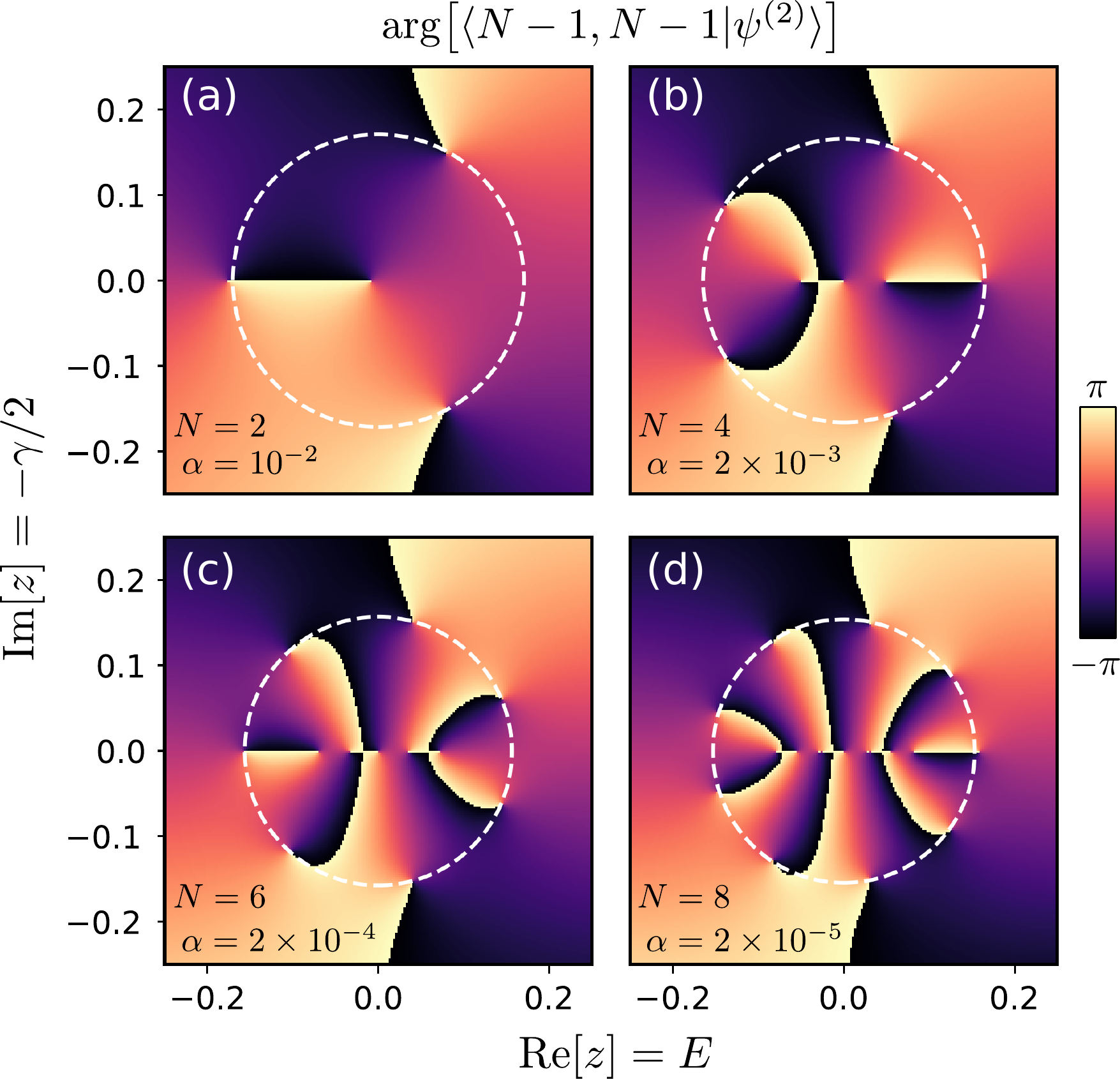}
  \caption{Complex argument of $\langle N-1,N-1|\psi^{(2)}\rangle$, the two-photon amplitude in the signal resonator, plotted versus $\mathrm{Re}[z] = E$ and $\mathrm{Im}[z] = -\gamma/2$.  The heat maps are obtained by solving the weak drive equations numerically, without the analytic approximations, using intracell coupling $t=0.1$ and different values of $N$, $\alpha$: (a) $N=2$ with $\alpha = 10^{-2}$, (b) $N=4$ with $\alpha=2\times10^{-3}$, (c) $N=6$ with $\alpha=2\times10^{-4}$, and (d) $N=8$ with $\alpha=2\times10^{-5}$.  The dashed circles indicate the optimal values of $|z|$ predicted by the analytic expression~\eqref{eq:alpha_z_final}.}
  \label{fig:wdl}
\end{figure}

\begin{figure*}
  \includegraphics[width=0.99\linewidth]{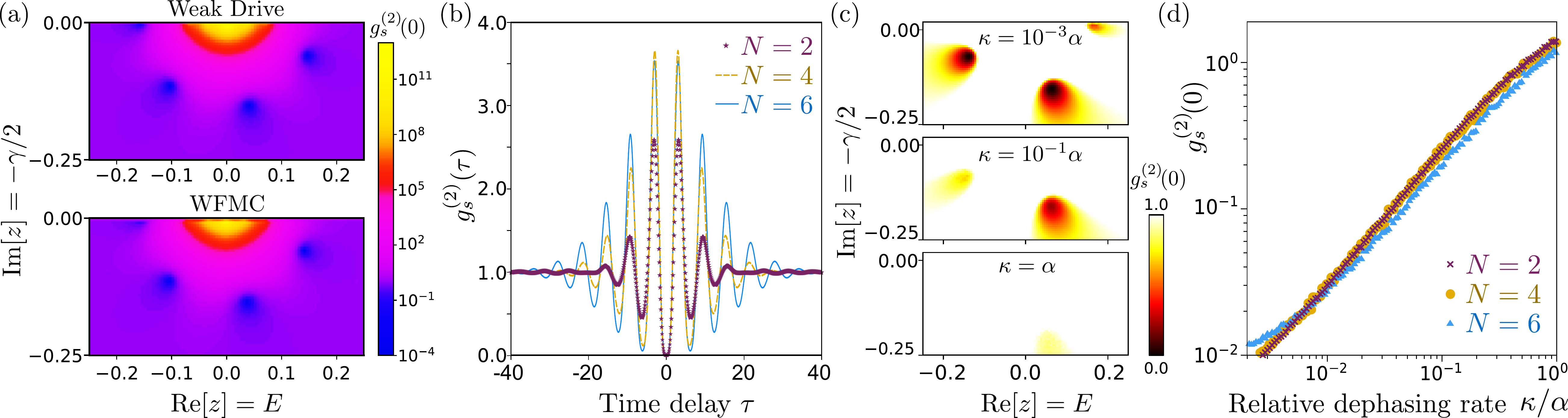}
  \caption{Wavefunction Monte Carlo (WFMC) results. In all subplots, we take $t = 0.1$ and for lattices of size $N=2$, $4$, $6$, we set $\alpha=10^{-2}$, $2\times10^{-3}$, $2\times10^{-4}$ respectively.  (a) Equal time second order correlation $g_s^{(2)}(0)$ versus $z = E - i\gamma/2$ for a lattice of size $N = 6$, obtained using the weak drive equations without series expansions (upper panel), and using WFMC simulations with drive amplitude $F_1 = 10^{-4}$ (lower panel).  (b) WFMC results for the unequal time correlation $g_s^{(2)}(\tau)$.  For each $N$, the value of $z$ is chosen to minimize $g_s^{(2)}(0)$, and is located by doing simulations with different $z$ on a discrete grid.  The horizontal axis has units of $\hbar$ over the intracell coupling.  In (a) and (b), the WFMC simulations do not include pure dephasing. (c) Plots of $g_s^{(2)}(0)$ versus $z$ for a lattice with $N=4$, obtained using WFMC simulations with dephasing rates $\kappa=0.001\alpha$, $0.1\alpha$, $\alpha$ (top to bottom).  To distinguish the local minima more clearly, all regions for which $g^{(2)}_s(0)\geq1$ are colored white. (d) Dependence of $g_s^{(2)}(0)$ on $\kappa/\alpha$ for lattices of size $N=2$, $4$, $6$, using the same values of $z$ as in (b). }
  \label{fig:wfmc}
\end{figure*}

Since $\alpha \in \mathbb{R}$, $z / |z|$ must be one of the $(N+1)$ complex roots of $(-1)^{N/2}$ or $(-1)^{N/2+1}$ depending on the sign of $\alpha$. We require $\mathrm{Im}(z) = -\gamma/2 < 0$ (i.e., the resonators are subject to loss rather than gain), and select the root with the most negative imaginary part.  This corresponds to the experimentally preferred situation where $\alpha/\gamma$, the nonlinearity strength relative to the cavity decay rate, is minimal.  (Choosing a different root yields only small modifications to the following results.) Here and in the following, we restrict our discussion to $\alpha>0$. Note that our theory also applies for $\alpha<0$. The results for $E$ and $\alpha$, expressed in terms of $\gamma$, are
\begin{align}
  E &= \frac{\gamma}{2} \cot{\theta},\\ 
  \alpha &= \frac{1}{4}\frac{(N-2){!!}}{(N-3){!!}}\big(\gamma  \csc\theta\big)^{N+1}, \label{eq:alpha_scaling} \\
  \theta &= \frac{N}{N+1}\, \frac{\pi}{2}.
\end{align}
For $N=2$, this reproduces the previously-derived single-dimer result \cite{Bamba11}.  For larger $N$, given any decay rate $\gamma$ smaller than the intercell coupling strength (i.e., $\gamma < 1$), Eq.~\eqref{eq:alpha_scaling} states that the nonlinearity strength $\alpha$ required for UPB decreases exponentially with the lattice size $N$.  This is the primary finding of the present work.

%% In Fig.~\ref{fig:wdl}, we draw dashed circles centered at the origin with radius $|z|$ calculated from Eq.(\ref{eq:alpha_z_final}) for each lattice to compare with numerical result.

Instead of using the series expansions, we can also solve the single-photon eigenproblem \eqref{eq:spectrum} numerically, and plug the results into the weak drive equations \eqref{eq:psi1} and \eqref{eq:psi2_0}--\eqref{eq:psi2_1} to obtain the single-photon and two-photon wavefunctions.  This is more efficient than directly solving Eqs.~\eqref{eq:steady1}--\eqref{eq:steady2} since it avoids performing a matrix inversion for each $z$.  Fig.~\ref{fig:lattice}(b) plots the optimal Kerr coefficients versus lattice size $N$, as given by the analytic expression derived from the series expansions (solid curves) and the numerical solutions to the weak drive equations (discrete points).  The analytic expression agrees well with the numerical results.

It is worth noting that the vanishing of the two-photon signal $\langle N-1, N-1|\psi^{(2)}\rangle$ is due to the presence of exact zeros in the complex $z$ plane.  It is a feature of the weak drive equations themselves, and is not an artifact of the series expansions leading up to the analytic result \eqref{eq:alpha_scaling}, which merely identify approximate (but accurate) locations for the zeros.  To show this, Fig.~\ref{fig:wdl} plots the complex argument of $\langle N-1,N-1|\psi^{(2)}\rangle$ versus $z$, obtained numerically from the weak drive equations for different $N$.  Here and in the following numerical examples, we take $t=0.1$ and choose different values of $\alpha$ for each $N$.  We observe phase singularities at discrete $z$ points, corresponding to analytic zeros.  These occur in the predicted pattern of roots of $(-1)^{N/2}$, and with increasing $N$ we see that smaller values of $\alpha$ are required for the zeros to appear at comparable $|z|$, as expected.

To verify the results obtained from the weak drive approximation, we performed WFMC simulations \cite{Dum92, Molmer93, Carmichael_1993, Barchielli_1991}, which solve the Lindblad question \eqref{eq:Lindblad} including the effects of stochastic quantum jumps.  Fig.~\ref{fig:wfmc}(a) compares the equal time second order correlations calculated by the two methods, for a lattice of $N = 6$ sites.  The WFMC simulations were performed using per-site Fock cutoffs chosen to balance accuracy and computational cost \cite{SM}; the results shown here were obtained with drive amplitude $F_1 = 10^{-4}$, but almost identical outcomes can be obtained for other values of $F_1 \ll 1$ \cite{SM}.  The WFMC and weak drive calculations produce very similar results, particularly with regard to the parameter values where photon antibunching occurs.  The results for lattices of size $N=2$ and $N=4$ show similar good agreement, as shown in the Supplemental Materials \cite{SM}.

Fig.~\ref{fig:wfmc}(b) shows the unequal time second order correlation $g^{(2)}_s(\tau)$ \cite{SM}, calculated with WFMC for lattices of size $N=2, 4,6$ (each lattice is individually tuned to its optimal point).  For $N = 2$ (the case of a single dimer), the correlation has previously been shown to oscillate with $\tau$ \cite{Liew10}.  The behavior for larger values of $N$ is similar, with the oscillation frequency not significantly influenced by $N$. 

In systems subject to significant environmental perturbations, UPB may be constrained by pure dephasing, distinct from the dissipation-induced quantum jumps considered thus far.  To investigate this, we performed a set of WFMC simulations with the term
\begin{equation}
  i\frac{\kappa}{2}\sum_{j}
  (2\ad_j\a_j \rho\ad_j\a_j - \ad_j\a_j\ad_j\a_j\rho - \rho\ad_j\a_j\ad_j\a_j)
\end{equation}
added to the Lindblad master equation \cite{Liew10}.  Fig.~\ref{fig:wfmc}(c) shows the $z$-dependance of $g_s^{(2)}(0)$ for $N=4$ using the dephasing rates $\kappa=10^{-3}\alpha$, $0.1\alpha$, and $\alpha$.  We see that pure dephasing, if excessively strong, ``smears out'' the zeros of $g_s^{(2)}(0)$.  In Fig.~\ref{fig:wfmc}(d), we plot the dependence of $g_s^{(2)}(0)$ on the dephasing rate for lattices of size $N=2$, $4$, and $6$.  These results indicate that photon antibunching requires $\kappa$ to be small compared to $\alpha$.  This will have to be taken into account to achieve UPB in large-$N$ lattices with extremely weak nonlinearities.

%% \begin{figure}
%%     \includegraphics[width=0.95\linewidth]{dephasing_N4.pdf}
%%     \caption{\label{fig:dephasing}
%%     (a) WFMC result for $N=4$ case with varying dephasing rates $\kappa=0,0.01\alpha,0.1\alpha$ and $\alpha$ from top to bottom, blank region represents $g^{(2)}_s\geq1$. (b) $g_s^{(2)}(0)$ at optimum $z$ with varying ratio of dephasing rate to nonlinearity strength for $N=2,4,6$ lattices.}
%% \end{figure}

Our finding of exponential enhancement of UPB in a dimer chain, one of the simplest lattice models, points to exciting opportunities for using photonic lattices to manipulate multi-photon quantum states.  In the future, related effects could be explored in more complicated systems such as two-dimensional lattices, as well as exploiting special lattice phenomena such as topologically protected single-photon states \cite{Wang09, Ozawa19, Peano2016, Wang2019}.  It would also be interesting to use these ideas to implement single-photon sources using silicon photonics, or other photonic platforms with weak optical nonlinearities.

This work was supported by Singapore MOE Academic Research Fund Tier~3 Grant MOE2016-T3-1-006, Tier~1 Grants RG187/18(S) and RG148/20, and Tier 2 Grants MOE2019-T2-2-085, MOE2019-T2-1-004

\bibliography{ref.bib}

\clearpage

\begin{widetext}

\appendix

\makeatletter 
\renewcommand{\theequation}{S\arabic{equation}}
\makeatother
\setcounter{equation}{0}

\makeatletter 
\renewcommand{\thefigure}{S\@arabic\c@figure}
\makeatother
\setcounter{figure}{0}

\makeatletter 
\renewcommand{\thesection}{S\arabic{section}}
\makeatother
\setcounter{section}{0}

\begin{center}
  {\Large Supplemental Materials}
\end{center}

\noindent

In these Supplemental Materials, we provide details of the analytic
derivations based on the weak drive approximation, as well as details
about the wave function Monte Carlo (WFMC) simulations.

\section{Weak drive approximation}

The few-photon quantum states of a system of driven coupled nonlinear optical cavities can be calculated using the weak drive approximation, a method developed in earlier works \cite{Bamba11,Flayac13,Flayac17}.  The system's Hibert space is divided into subspaces of different photon numbers, so that
\begin{align}
  \ket{\psi} &= |\psi^{(0)}\rangle + |\psi^{(1)}\rangle
  + |\psi^{(2)}\rangle + \cdots,\\
  |\psi^{(k)}\rangle &=
  \sum_{n_1+\hdots+n_N=k}c_{n_1n_2\hdots n_N}\ket{n_1,n_2,\dots, n_N},
\end{align}
where $|n_1,n_2,\cdots, n_N\rangle$ denotes the bosonic state with $n_1$ photons on site 1, $n_2$ photons on site 2, etc.  The vacuum state is $|\psi^{(0)}\rangle = \ket{0,\dots,0}$.

Amongst the various Hamiltonian terms discussed in the main text, $\Hami_c$, $\Hami_p^{\prime}$, and $\Hami_{nl}$ all conserve photon number, but the driving Hamiltonian $\Hami_d$ does not.  Let us split $\Hami_d$ as follows:
\begin{align}
  \Hami_d &= \Hami_++\Hami_-, \\
  \Hami_+ &= \sum_{j=1}^NF_j \ad_j \;\; = \Hami_-^\dagger.
\end{align}
In the weak drive limit, the amplitudes for higher photon number states are of sub-leading order, so
\begin{equation}
  \big\langle \psi^{(k)} \big| \Hami_- \big| \psi^{(k+1)} \big\rangle
  \ll \big\langle \psi^{(k)} \big| \Hami_+ \big| \psi^{(k-1)} \big\rangle.
\end{equation}
By matching particle number subspaces in the Schr\"odinger equation, we obtain
\begin{align}
  i\hbar \frac{d}{dt} |\psi^{(k)}\rangle
  &= \Hami |\psi^{(k)}\rangle
  + \Hami_+ |\psi^{(k-1)}\rangle
  + \Hami_- |\psi^{(k+1)}\rangle \\
  &\approx \Hami |\psi^{(k)}\rangle + \Hami_+ |\psi^{(k-1)}\rangle.
\end{align}
Truncating at $k=2$ (i.e., neglecting states with three or more photons) gives
\begin{empheq}[left=\empheqlbrace\;\;]{align}
  i\hbar\frac{d}{dt} |\psi^{(1)}\rangle
  &= \Hami |\psi^{(1)}\rangle + \Hami_+ |\psi^{(0)}\rangle, \\
  i\hbar \frac{d}{dt} |\psi^{(2)}\rangle
  &= \Hami |\psi^{(2)}\rangle + \Hami_+ |\psi^{(1)}\rangle.
\end{empheq}
We look for steady state solutions, for which the time derivatives are zero.  Hence,
\begin{empheq}[left=\empheqlbrace\;\;]{align}
  |\psi^{(1)}\rangle
  &=-\Hami^{-1} \Hami_{+} |\psi^{(0)}\rangle\label{SMeq:steady_1}\\
  |\psi^{(2)}\rangle
  &=-\Hami^{-1} \Hami_{+} |\psi^{(1)}\rangle. \label{SMeq:steady_2}
\end{empheq}

\section{Single-photon states}

We can solve Eq.~\eqref{SMeq:steady_1} by using the lattice's eigenmode basis.  Let $|j\rangle \equiv \ad_j\ket{\psi^{(0)}}$ be the state in which only site $j$ is occupied, with a single photon.  Projecting into the single-photon subspace and using these states as the basis, the Hamiltonian is
\begin{align}
  \Hami^{(1)} &= z\mathcal{I} + \Hami^{(1)}_c, \;\;\;\mathrm{where}\\
  z &= E-i\frac{\gamma}{2}, \qquad
  \Hami^{(1)}_c
  = \left(\sum_{j=1}^{N/2}\ket{2j-1}\bra{2j}
  +t\sum_{j=1}^{N/2-1}\ket{2j}\bra{2j+1}\right) + \mathrm{h.c.},
  \label{SMeq:Hc1}
\end{align}
with $\mathcal{I}$ denoting the identity operator.  Note that $\Hami_{nl}$ does not contribute at the single-photon level.

Denote the eigenstates of the single particle Hamiltonian $\Hami_c^{(1)}$ by $\{|\varphi_n\rangle\}$, with the corresponding energy eigenvalues $\{\epsilon_n\}$.  Using these eigenstates as an orthogonal basis for the single-particle Hibert space, we obtain
\begin{equation}
  \Hami^{(1)}\ket{\varphi_n}
  = \left(z+\epsilon_n\right)\ket{\varphi_n}.
\end{equation}
Hence, the resolvent operator (or Green's function) can be written as
\begin{equation}
    \Green^{(1)} \equiv \left(z\mathcal{I} + \Hami_c^{(1)}\right)^{-1}
  = \sum_n \frac{\ket{\varphi_n}\bra{\varphi_n}}{z+\epsilon_n}.
  \label{eq:green1}
\end{equation}
Plugging this into Eq.~\eqref{SMeq:steady_1} yields
\begin{align}
  |\psi^{(1)}\rangle
  &= -\sum_{j=1}^N \,F_j \, \Green^{(1)} |j\rangle
  \label{SMeq:psi1_green} \\
  &=-\sum_n \frac{f_n\ket{\varphi_n}}{z+\epsilon_n},
  \;\;\; \mathrm{where}\;\; f_n=\sum_{j=1}^N F_j \braket{\varphi_n}{j}.
  \label{SMeq:psi1}
\end{align}

\section{Two-photon states: linear limit}

Next, we deal with the two-photon part of the weak drive solution, which is given by Eq.~\eqref{SMeq:steady_2}.  The two-particle subspace can be spanned using tensor product states such as $\ket{i,j}=\ket{i}\otimes\ket{j}$ (site basis) or $\ket{\varphi_{kl}}=\ket{\varphi_k}\otimes\ket{\varphi_l}$ (single-particle eigenbasis).  Note that the basis vectors are not symmetrized, but the two-photon states themselves should be symmetrized, e.g.~$\braket{i,j}{\psi^{(2)}}=\braket{j,i}{\psi^{(2)}}$.

Using Eq.~\eqref{SMeq:psi1_green},
\begin{align}
  \mathcal{H}_+ |\psi^{(1)}\rangle 
  &=-\mathcal{H}_+ \sum_{j,j'} \,F_j \, |j'\rangle
  \langle j'| \Green^{(1)} |j\rangle \\
  &=-\sum_{i,j, j'} \Big(F_i \,a_i^\dagger \Big)
  \Big( F_j\, a_{j'}^\dagger |\psi^{(0)}\rangle
  \langle j'|\Green^{(1)} |j\rangle \Big) \\
  &=-\frac{1}{\sqrt{2}} \sum_{i,j} F_i F_j
  \left(|i\rangle \otimes \Green^{(1)}|j\rangle
  + \Green^{(1)}|j\rangle \otimes |i\rangle \right).
  \label{eq:resummedHplupsi}
\end{align}
In deriving Eq.~\eqref{eq:resummedHplupsi}, we are careful to distinguish same-site and different-site excitations: $a^\dagger_ia^\dagger_i|\psi^{(0)}\rangle = \sqrt{2} |i,i\rangle$, whereas 
$a^\dagger_ia^\dagger_j|\psi^{(0)}\rangle = \frac{1}{\sqrt{2}}\big( |i,j\rangle + |j,i\rangle\big)$ for $i\ne j$.  Hence, in the eigenstate basis,
\begin{equation}
  \mathcal{H}_+ |\psi^{(1)}\rangle
  =-\frac{1}{\sqrt{2}} \sum_{mn} f_mf_n\,
  \frac{2z+\epsilon_m+\epsilon_n}{(z + \epsilon_m)(z + \epsilon_n)} \;
  |\varphi_{mn}\rangle. \label{SMeq:s2}
\end{equation}

Next, let us take the linear limit for the moment (i.e., neglecting $\Hami_{nl}$).  The projection of the Hamiltonian into the two-particle subspace is
\begin{equation}
  \Hami_0^{(2)}
  =\Hami^{(1)}\otimes\mathcal{I} + \mathcal{I}\otimes\Hami^{(1)},
\end{equation}
and so, in the basis formed by the eigenstates,
\begin{align}
  \Hami_0^{(2)}\ket{\varphi_{mn}}
  &=\left(\Hami^{(1)} \otimes \mathcal{I}
  + \mathcal{I} \otimes \Hami^{(1)}\right)
  \ket{\varphi_m}\otimes\ket{\varphi_n} \notag\\
  & =(2z+\epsilon_m+\epsilon_n)\ket{\varphi_{mn}}.
\end{align}
Hence, the two-photon Green's function, in the linear limit, is
\begin{equation}
  \Green_0^{(2)} \equiv \left(\Hami_0^{(2)}\right)^{-1}
  = \sum_{mn} \frac{\ket{\varphi_{mn}}\bra{\varphi_{mn}}}{2z + \epsilon_m + \epsilon_n}.
  \label{SMeq:h2_0_inv}
\end{equation}
Plugging Eqs.~\eqref{SMeq:h2_0_inv} and \eqref{SMeq:s2} into Eq.~\eqref{SMeq:steady_2}, we obtain the two-photon state in the linear limit: 
\begin{align}
  |\psi_0^{(2)}\rangle &= \frac{1}{\sqrt{2}}
  \sum_{mn}\frac{f_mf_n}{(z+\epsilon_m)(z+\epsilon_n)}\ket{\varphi_{mn}}
  \label{SMeq:psi2_0_0} \\
  &= \frac{1}{\sqrt{2}}
  \left(\sum_m\frac{f_m\ket{\varphi_m}}{z+\epsilon_m}\right)
  \otimes \left(\sum_n\frac{f_l\ket{\varphi_n}}{z+\epsilon_n}\right) \\
  &= \frac{1}{\sqrt{2}} |\psi^{(1)}\rangle \otimes |\psi^{(1)}\rangle.
  \label{SMeq:psi2_0_factorization}
\end{align}

Hence, in the linear limit, the the zero delay time second order correlation functions are unity, as expected:
\begin{equation}
    g^{(2)}_{ij}(0)\Big|_{\alpha=0}=2\left|\frac{\braket{i,j}{\psi_0^{(2)}}}{\braket{i}{\psi^{(1)}}\braket{j}{\psi^{(1)}}}\right|^2=1.
\end{equation}

\section{Two-photon states: perturbative correction}

We now have to deal with the nonlinear term
\begin{equation}
  \Hami^{(2)}_{nl}=2\alpha\sum_i\ket{i,i} \bra{i,i} \equiv\alpha \V.
  \label{eq:Vdef}
\end{equation}
For small $\alpha$, this can be treated perturbatively using the Dyson series:
\begin{align}
  \Green^{(2)} \equiv \left(\Hami^{(2)}\right)^{-1}
  = \left(\Hami_0^{(2)}+\alpha \V\right)^{-1}
  = \Green^{(2)}_0 - \alpha \;\Green_0^{(2)} \, \V\, \Green_0^{(2)} + \cdots
  \label{SMeq:perturbation}
\end{align}
Truncating away the terms denoted by ellipses, and plugging back into Eq.~\eqref{SMeq:steady_2}, gives the result
\begin{align}
  |\psi^{(2)}\rangle &\approx |\psi^{(2)}_0\rangle
  + \alpha\, |\psi^{(2)}_1\rangle,
  \label{SMeq:psi2} \\
  |\psi_0^{(2)}\rangle &= \frac{1}{\sqrt{2}} |\psi^{(1)}\rangle \otimes |\psi^{(1)}\rangle
  \label{SMeq:psi2_0} \\
  |\psi^{(2)}_1\rangle
  %% &= \Green^{(2)}_0 \, \V \, \Green^{(2)}_0 \, \Hami_+ |\psi^{(1)} \rangle
  %% \notag\\
  &= \sum_{imnpq} \frac{-\sqrt{2}f_{p}f_{q} \braket{\varphi_{mn}}{i,i} \braket{i,i}{\varphi_{pq}} }{(z+\epsilon_{p}) (z+\epsilon_{q})(2z+\epsilon_m+\epsilon_n)}
  \; |\varphi_{mn}\rangle. \label{SMeq:psi2_1} \\
  &= -\sqrt{2} \sum_{ijj'} F_j F_{j'} \Green_0^{(2)} |i,i\rangle \langle i|
  \Green^{(1)} |j\rangle \langle i|\Green^{(1)} |j'\rangle.
%                      &=-\sqrt{2}\sum_{p,q}F_pF_q\sum_{k,l}\left(\sum_i\bra{i,i}\sum_{k',l'}\frac{\ket{\varphi_{k' l'}}\bra{\varphi_{k' l'}}}{(z+\epsilon_{k'})(z+\epsilon_{l'})}\ket{p,q}\braket{\varphi_{kl}}{i,i}\right)\frac{\ket{\varphi_{kl}}}{2z+\epsilon_k+\epsilon_l}.
\end{align}
In deriving Eq.~\eqref{SMeq:psi2_1}, we have combined Eqs.~\eqref{SMeq:h2_0_inv}, \eqref{SMeq:s2}, and \eqref{eq:Vdef}.

%% \section{Role of chiral symmetry for $N=2$ case}
%% For the single dimer lattice, the eigenenergies and eigenstates of $\Hami_c^{(1)}$ are:
%% \begin{equation}
%% 	\epsilon_1=1, \ket{\varphi_1}=\frac{1}{\sqrt{2}}\begin{pmatrix}1\\1\end{pmatrix}; \epsilon_2=-1,\ket{\varphi_2}=\frac{1}{\sqrt{2}}\begin{pmatrix}1\\-1\end{pmatrix}.
%% \end{equation}
%% In Eq.(\ref{SMeq:psi2_1}), when $|z|$ is small, only $\ket{\varphi_{12}}$ and $\ket{\varphi_{21}}$ has prominent contribution. Consider the case where only the first cavity is driven, Eq.(\ref{SMeq:psi2_1}) gives
%% \begin{equation}
%% 	\ket{\psi_1^{(2)}}\approx\frac{F_1^2}{2z}\left(\frac{\ket{\varphi_{12}}+\ket{\varphi_{21}}}{\sqrt{2}}\right).
%% \end{equation}
%% The first order perturbation of the two-photon wavefunction is proportional to the zero-energy two-photon eigenstate $(\ket{\varphi_{12}}+\ket{\varphi_{21}})/\sqrt{2}$, which is the tensor product of a single-photon eigenstate with its chiral symmetric partner.

\section{Matrix elements of the single-photon Green's function}
\label{sec:Hc_property}

In Eq.~\eqref{eq:green1}, we expressed the single-photon resolvent \eqref{eq:green1} using the eigenbasis of the single-particle SSH Hamiltonian $\Hami_c^{(1)}$, defined in Eq.~\eqref{SMeq:Hc1}.  Within the disk $|z|<1$, it can be further expanded as a Laurent series, as follows:
\begin{align}
  \Green^{(1)} = \sum_n \frac{\ket{\varphi_n}\bra{\varphi_n}}{z+\epsilon_n}
  &= \sum_n \epsilon_n^{-1} \left(1+\frac{z}{\epsilon_n}\right)^{-1}
  \ket{\varphi_n}\bra{\varphi_n} \\
  &= \sum_n \left(\epsilon_n^{-1}-\epsilon_n^{-2}z
  + \epsilon_n^{-3}z^2 - \cdots\right) \ket{\varphi_n}\bra{\varphi_n} \\
  &= g_c  - z g_c^2 + z^2 g_c^3  - \cdots,
  \label{eq:green1_expansion}
\end{align}
where
\begin{equation}
  g_c \equiv \left(\Hami_c^{(1)}\right)^{-1}.
  \label{eq:gc}
\end{equation}
%
%% When $t\ll1$, the Hamiltonian exhibits two symmetric bands centered at $\epsilon=\pm 1$, with narrow bandwidth $2t$.  Due to the chiral symmetry of the Hamiltonian, the eigenvalues appear in pairs of opposite signs.
%
We will need the matrix elements of $g_c$.  For this, it is useful to write the SSH Hamiltonian in matrix form as
\begin{equation}
  \Hami_c^{(1)}=\begin{bmatrix}
  0&1& & & \\
  1&0&t& & \\
  &t&0&1& \\
  & &1&0&\sddots\\
  & & &\sddots&\sddots
  \end{bmatrix}_{N\times N}=H_0+tH_1, 
  \label{eq:SSH_decomp}
\end{equation}
where
\begin{equation}
  H_0=\begin{bmatrix}A&&&\\&A&&\\&&A&\\&&&\sddots\end{bmatrix}, \;\;
  H_1=\begin{bmatrix}0&B&&\\B^T&0&B&\\&B^T&0&\sddots\\&&\sddots&\sddots\end{bmatrix}, \;\;
  A=\begin{bmatrix}0&1\\1&0\end{bmatrix}, \;\;
  B=\begin{bmatrix}0&0\\1&0\end{bmatrix}.
  \label{eq:H0etc}
\end{equation}
For brevity, we denote
\begin{equation}
  U=AB=\begin{bmatrix}1&0\\0&0\end{bmatrix},\quad
  D=AB^T=\begin{bmatrix}0&0\\0&1\end{bmatrix}.
\end{equation}
These matrices have the following easily-verified properties:
\begin{align}
  H_0 &=H_0^{-1} \label{eq:H0inv} \\
  D^2 &=D \\
  U^2 &=U \\
  UD=DU &=0.
\end{align}
Moreover,
\begin{equation}
  H_0H_1 = \begin{bmatrix}
    0&U&&&\\D&0&U&&\\&D&0&U&\\&&D&0&\sddots\\&&&\sddots&\sddots
  \end{bmatrix}, \quad 
  (H_0H_1)^2 = \begin{bmatrix}
    0&0&U&&\\0&0&0&U&\\D&0&0&0&\sddots\\&D&0&0&\sddots\\
    &&\sddots&\sddots&\sddots
  \end{bmatrix} \quad\dots\quad
  (H_0H_1)^{N/2-1} = \begin{bmatrix}
    0&0&\shdots&0&U\\0&0&&&0\\\svdots&&\sddots&&\svdots\\
    0&&&0&0\\D&0&\shdots&0&0
  \end{bmatrix}. \label{eq:h0h1_power}
\end{equation}
For any $n>N/2-1$, $(H_0H_1)^n=0$.  

We now seek the following matrix element for each of the terms in the Laurent series \eqref{eq:green1_expansion}:
\begin{equation*}
  \langle N-1 | \,g_c^{m+1} | 1 \rangle, \;\;\; m = 0, 1, 2, \dots,
\end{equation*}
with $g_c$ defined in Eq.~\eqref{eq:gc}.  These are the matrix elements connecting the driven site, $j=1$, and the signal site, $j = N-1$.  We will work in the small $t$ regime, and look for solutions to lowest non-vanishing order in $t$.  Using Eqs.~\eqref{eq:SSH_decomp} and \eqref{eq:H0inv}, $g_c$ can be expressed as a series in $t$:
\begin{equation}
  g_c = \left(H_0+tH_1\right)^{-1} = H_0\sum_{n=0}^{\infty}(-tH_0H_1)^n.
  \label{eq:greenc_expansion}
\end{equation}
Now, for the zeroth-order term in the Laurent series,
\begin{equation}
  \bra{N-1} \, g_c \ket{1}
  = \sum_{n=0}^{\infty} (-t)^n \; \bra{N-1} \, H_0\,(H_0H_1)^n \ket{1}.
  \label{eq:zeroth_order_g}
\end{equation}
Referring to Eq.~\eqref{eq:h0h1_power}, all terms with $n < N/2 - 1$ have vanishing matrix element (roughly speaking, there are not enough ``powers of $t$'' to propagate from the first unit cell to the last unit cell).  As for the next term, which is of order $n = N/2 - 1$,
\begin{align}
  \bra{N-1} \, H_0\,(H_0H_1)^{N/2-1} \ket{1}
  &= \;
  \raisebox{6.3mm}{$\begin{bmatrix}0 & \cdots & 0 & \bra{+z} \end{bmatrix}$}
  \begin{bmatrix}A&&&\\&A&&\\&&\sddots&\\&&&A\end{bmatrix}
    \begin{bmatrix}
      0&\shdots&0&U\\
      0&&&0\\
      \svdots&\sddots&&\svdots\\
      D&\shdots&0&0
    \end{bmatrix}
    \begin{bmatrix}\ket{+z} \\ 0 \\ \svdots \\ 0 \end{bmatrix}
    \;\;\;\mathrm{where}\;\; \ket{+z} \equiv 
    \begin{bmatrix}
      1 \\ 0
    \end{bmatrix}
    \label{eq:hairy1} \\
  &= \;\raisebox{6.2mm}{$\begin{bmatrix}0 & \cdots & 0 & \bra{+z}
  \end{bmatrix}$}
    \begin{bmatrix}0 \\ \svdots\\ 0 \\ AD\ket{+z}
  \end{bmatrix} \\
  &= \;\raisebox{2.2mm}{$\begin{bmatrix}1 & 0 \end{bmatrix}$}
  \begin{bmatrix}
    0 & 1 \\ 0 & 0
  \end{bmatrix}
  \begin{bmatrix}
    1 \\ 0
  \end{bmatrix}
  \;=\; 0.
  \label{eq:hairy3}
\end{align}
Hence,
\begin{equation}
  \bra{N-1} \, g_c \ket{1} \sim O\left(t^{\frac{N}{2}}\right).
  \label{eq:gcorder}
\end{equation}

We now proceed to the $O(z^1)$ term in the Laurent series.  For this, we need to consider the double series
\begin{equation}
  g_c^2
  = \left(H_0\sum_{n=0}^{\infty}(-tH_0H_1)^n\right)^2
  =\sum_{n=0}^{\infty}(-t)^n\sum_{p=0}^n H_0(H_0H_1)^pH_0(H_0H_1)^{n-p}.
\end{equation}
Referring again to Eq.~\eqref{eq:h0h1_power}, the terms with $n < N/2 - 1$ have vanishing $(N-1, 1)$ matrix element.  The matrix element for the next order term is
\begin{equation*}
  (-t)^{\frac{N}{2}-1} \sum_{p=0}^{N/2-1}
  \bra{N-1} H_0(H_0H_1)^p H_0 (H_0H_1)^{\frac{N}{2}-1-p} \ket{1}.
\end{equation*}
For each $p$, we can evaluate the matrix expressions similarly to Eqs~\eqref{eq:hairy1}--\eqref{eq:hairy3}.  Because $ADAD = 0$, it turns out that the only non-vanishing term is $p = N/2-1$:
\begin{align}
  \bra{N-1} H_0 (H_0H_1)^{\frac{N}{2}-1} H_0 \ket{1} &=
  \bra{+z} ADA \ket{+z} = 1 \\
  \Rightarrow \;\;\;
  \langle N-1 | \,g_c^2\, | 1\rangle
  &= (-t)^{\frac{N}{2}-1} + O\left(t^{\frac{N}{2}}\right).
\end{align}
Notably, the leading term is lower order in $t$ than the contribution from the zeroth-order Laurent series term, Eq.~\eqref{eq:gcorder}.

It can similarly be seen that the higher order terms in the Laurent series,
involving $g_c^3$, $g_c^4$, and so forth, all have vanishing contributions
to order $t^{N/2-1}$ or below.  Thus,
\begin{equation}
  \langle N-1 | \Green^{(1)} | 1\rangle
  = - z (-t)^{\frac{N}{2}-1} + O\left(t^{\frac{N}{2}}\right).
  \label{eq:Green1_result}
\end{equation}
This expression is valid to lowest order in $t$, but to \textit{all}
orders in $z$ within the circle of convergence $|z| < 1$.

\section{Photon blockade condition}

We now return to the two-photon wavefunctions, whose two leading terms are given by Eqs.~\eqref{SMeq:psi2}--\eqref{SMeq:psi2_1}.

First, consider the zeroth-order term \eqref{SMeq:psi2_0}, and consider the two-photon amplitude in the signal resonator, under the driving condition $f_n = F_1 \langle\varphi_n|1\rangle$.  From Eqs.~\eqref{SMeq:psi1_green} and \eqref{SMeq:psi2_0_factorization},
\begin{align}
  \langle N-1,N-1 | \psi_0^{(2)}\rangle
  &= \frac{1}{\sqrt{2}}
  \langle N-1, N-1 | \Big(|\psi^{(1)}\rangle \otimes |\psi^{(1)}\rangle\Big) \\
  &= \frac{F_1^2}{\sqrt{2}} \Big( \langle N-1 | \Green^{(1)} |1\rangle \Big)^2
\end{align}
Hence, using Eq.~\eqref{eq:Green1_result},
\begin{equation}
  \langle N-1,N-1 | \psi_0^{(2)}\rangle \,\approx\,
  \frac{F_1^2}{\sqrt{2}} \;z^2\, t^{N-2}.
  \label{SMeq:z2}
\end{equation}
%
%% \begin{equation}
%%     \bra{i,i}\sum_{k,l}\frac{\ket{\varphi_{kl}}\bra{\varphi_{kl}}}{(z+\epsilon_k)(z+\epsilon_l)}\ket{1,1}\approx\delta_{i2},\label{SMeq:z0}
%% \end{equation}
%

Next, we turn to the first-order term \eqref{SMeq:psi2_1}.  Before taking the matrix elements, let us perform a few simplifications:
\begin{align}
  |\psi_1^{(2)}\rangle
  &= \sum_{imnpq} \ket{\varphi_{mn}} \frac{-\sqrt{2}f_{p}f_{q} \braket{\varphi_{mn}}{i,i} \braket{i,i}{\varphi_{pq}} }{(z+\epsilon_{p})(z+\epsilon_{q})(2z+\epsilon_m+\epsilon_n)} \\
  &=-\sqrt{2}F_1^2 \sum_{imn} \bra{i,i} \sum_{pq}
  \frac{\ket{\varphi_{pq}}\bra{\varphi_{pq}}}{(z+\epsilon_{p})(z+\epsilon_{q})}
  \ket{1,1}
  \frac{\braket{\varphi_{mn}}{i,i}}{2z+\epsilon_m+\epsilon_n}\;
  \ket{\varphi_{mn}}\\
  &=-\sqrt{2}F_1^2 \sum_{imn} 
  \frac{\big[\bra{i} \Green^{(1)} \ket{1}\big]^2 \bra{\varphi_{mn}}
    \ket{i,i}}{2z+\epsilon_m+\epsilon_n}\; \ket{\varphi_{mn}}
\end{align}
The $\langle i | \Green^{(1)} | 1\rangle $ in the numerator can be evaluated to lowest order in $t$.  Referring back to Eqs.~\eqref{eq:green1_expansion} and \eqref{eq:greenc_expansion}, only the $i = 2$ case yields a zeroth order contribution:
\begin{equation}
  \bra{i}\Green^{(1)}\ket{1} = \bra{i} \left(g_c + \cdots\right) \ket{1}
  = \bra{i}H_0\ket{1}+ \cdots =\delta_{i2}+\cdots,
\end{equation}
where ellipses denote terms of $O(t)$ and higher.  Hence,
\begin{align}
  |\psi_1^{(2)}\rangle &\approx -\sqrt{2} F_1^2
  \sum_{mn} \frac{\ket{\varphi_{mn}}
    \braket{\varphi_{mn}}{2,2}}{2z+\epsilon_m+\epsilon_n} \\
  &= -\sqrt{2}F_1^2
  \left(\sum_{\epsilon_m\epsilon_n>0} + \sum_{\epsilon_m\epsilon_n<0}\right)
  \frac{\ket{\varphi_{mn}} \braket{\varphi_{mn}}{2,2}}
       {2z + \epsilon_m + \epsilon_n}
\end{align}
In the last line, we have split the sums over eigenstates to account for the fact that the single-particle SSH model has both positive- and negative-energy states.  The model's chiral symmetry ensures that the eigenstates occur in pairs with energies of opposite signs.  Evidently, the largest contributions in the above sums come from the terms with $\epsilon_m$ and $\epsilon_n$ of opposite signs, so that they can (partially or fully) cancel each other in the denominator:
\begin{equation}
  |\psi_1^{(2)}\rangle \approx -\sqrt{2}\, F_1^2 \sum_{\epsilon_m\epsilon_n<0}
  \frac{\ket{\varphi_{mn}} \braket{\varphi_{mn}}{2,2}}{2z + \epsilon_m + \epsilon_n}.
  \label{eq:psi12_intermediate}
\end{equation}
Referring to Eqs.~\eqref{eq:SSH_decomp}--\eqref{eq:H0etc}, we can write
\begin{align}
  \braket{\varphi_n}{2}=\bra{\varphi_n}H_0\ket{1}
  =\bra{\varphi_n} \left(\Hami_c^{(1)}-tH_1\right) \ket{1}
  =\epsilon_n\braket{\varphi_n}{1}.
  \label{eq:first_cell}
\end{align}
Plugging this into Eq.~\eqref{eq:psi12_intermediate} gives
\begin{align}
  |{\psi_1^{(2)}}\rangle
  &\approx -\sqrt{2}\, F_1^{2} \sum_{\epsilon_m\epsilon_n<0}
  \frac{\epsilon_m\epsilon_n\ket{\varphi_{mn}}\braket{\varphi_{mn}}{1,1}}{2z+\epsilon_m+\epsilon_n}.
  \label{eq:psi12_intermediate1}
\end{align}
Hence,
\begin{align}
  \langle N-1, N-1|\psi_1^{(2)}\rangle
  &= - \sqrt{2} \, F_1^2
  \left[\sum_{\epsilon_m < 0} \sum_{\epsilon_n>0}
    \frac{A_{mn}}{2z+\epsilon_m+\epsilon_n}
    + \sum_{\epsilon_m > 0} \sum_{\epsilon_n <0}
    \frac{A_{mn}}{2z+\epsilon_m+\epsilon_n}
    \right],
  \label{eq:psi12_intermediate3} \\
  A_{mn} &= \epsilon_m \epsilon_n
  \langle N-1 | \varphi_m\rangle  \;
  \langle \varphi_{m} | 1 \rangle \;
  \langle N-1 | \varphi_{n} \rangle \;
  \langle \varphi_{n} | 1 \rangle.
  \label{eq:Amn}
\end{align}
The chiral symmetry operator
\begin{equation}
  \Gamma=\begin{bmatrix}1&&&&\\&-1&&&\\&&1&&\\&&&-1&\\&&&&\sddots\end{bmatrix}_{N\times N}
\end{equation}
anticommutes with $\Hami_c^{(1)}$ and maps each eigenstate to one with an eigenvalue of the opposite sign:
\begin{equation}
  \Hami_c^{(1)}\, \Gamma\ket{\varphi_n} =-\epsilon_n \Gamma \ket{\varphi_n}
  \label{eq:chiral_symmetry}.
\end{equation}
Using this, we can replace the sum over the negative-energy states in Eq.~\eqref{eq:psi12_intermediate3} (say index $n$) with a sum over positive energy states, with the terms in the sum modified by replacing $|n\rangle$ with $\Gamma|n\rangle$ and $\epsilon_n$ with $-\epsilon_n$.  It is easily shown that the $A_{mn}$ factor switches sign in the process, so that
\begin{align}
  \langle N-1, N-1|\psi_1^{(2)}\rangle
  = \sqrt{2} \, F_1^2 \sum_{\epsilon_m,\epsilon_n>0}
  A_{mn} \left[\frac{1}{2z+(\epsilon_m-\epsilon_n)}
    + \frac{1}{2z-(\epsilon_m-\epsilon_n)}\right]
\end{align}
For an infinite dimer chain, the width of the upper band is $2t$. For a finite chain, we have $|\epsilon_m-\epsilon_n| \lesssim 2t$ for $\epsilon_m,\epsilon_n>0$. For $|z|>t$, we can take the Laurent expansion
\begin{align}
  \langle N-1, N-1|\psi_1^{(2)}\rangle
  &= 2 \sqrt{2} \, F_1^2 \sum_{\epsilon_m,\epsilon_n>0} \frac{A_{mn}}{2z}
  \sum_{p=0}^{\infty}\left(\frac{\epsilon_m-\epsilon_n}{2z}\right)^{2p} \\
  &=2\sqrt{2} \, F_1^2 \; \sum_{p=0}^{\infty}
  (2z)^{-(2p+1)}\sum_{\epsilon_m,\epsilon_n>0}A_{mn}(\epsilon_m-\epsilon_n)^{2p}
  \label{SMeq:psi2_1_second_nast}.
\end{align}
%% Here the Laurent expansion is done in the annulus $|z|>t$ since for the coupling Hamiltonian $\Hami_c^{(1)}$, its positive energy band's width is approximately $2t$, so $|\epsilon_k-\epsilon_l|<2t$ for any $\epsilon_k,\epsilon_l>0$. Note the difference between summation subscripts $\epsilon_k\epsilon_l$ and $\epsilon_k,\epsilon_l$.
The inner sum in Eq.~\eqref{SMeq:psi2_1_second_nast} can be evaluated with the help of Eq.~\eqref{eq:Amn}:
\begin{align}
  \sum_{\epsilon_m,\epsilon_n>0} A_{mn} (\epsilon_m-\epsilon_n)^{2p}
  &= \sum_{\epsilon_m,\, \epsilon_n>0} A_{mn}
  \sum_{h=0}^{2p} (-1)^h \binom{2p}{h} \epsilon_m^{2p-h} \epsilon_n^h
  \\
  &= \sum_{\epsilon_m,\epsilon_n>0}
  \langle N-1 | \varphi_m\rangle
  \langle \varphi_{m} | 1 \rangle
  \langle N-1 | \varphi_{n} \rangle
  \langle \varphi_{n} | 1 \rangle 
  \sum_{h=0}^{2p} (-1)^h \binom{2p}{h} \epsilon_m^{2p-h+1} \epsilon_n^{h+1}
  \\
  &= \sum_{h=0}^{2p} (-1)^h \binom{2p}{h}
  \langle N-1 | (\mathcal{H}_c^+)^{2p-h+1} | 1 \rangle \;
  \langle N-1 | (\mathcal{H}_c^+)^{h+1} | 1 \rangle,
  \label{eq:Amnsum_intermediate}
\end{align}
where
\begin{equation}
  \Hami_c^{+}=\sum_{\epsilon_n > 0} \epsilon_n \ket{\varphi_n} \bra{\varphi_n}.
\end{equation}
Similar to our earlier treatment of $\Hami_c^{(1)}$, we can show that
\begin{equation}
  \bra{N-1}(\Hami_c^+)^\nu \ket{1}
  = \frac{1}{2} \, G(N-2, \nu) \, t^{\frac{N}{2}-1} + O(t^{\frac{N}{2}}),
  \label{SMeq:Hcpn}
\end{equation}
where
\begin{equation}
  G(a,b)\equiv\frac{b{!}{!}}{a{!}{!}(b-a){!}{!}}.
\end{equation}
Applying this to Eq.~\eqref{eq:Amnsum_intermediate} yields
\begin{align}
  \sum_{\epsilon_m,\epsilon_n>0} A_{mn} (\epsilon_m-\epsilon_n)^{2p}
  &= \frac{t^{N-2}}{4}\, \sum_{h=0}^{2p} (-1)^h \binom{2p}{h}
  G(N-2, 2p-h+1) \, G(N-2, h+1) + O(t^{N-1}).
\end{align}

\begin{figure}
  \includegraphics[width=0.85\linewidth]{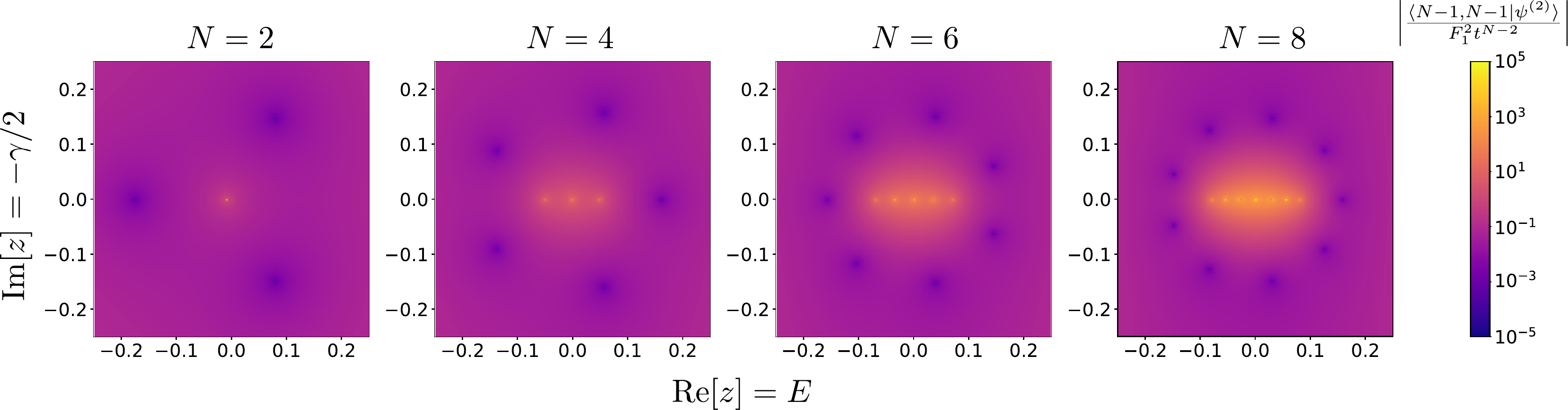}
  \caption{Normalized two-photon signal $\big|\langle N-1,N-1|\psi^{(2)}\rangle / (F_1^2t^{N-2})\big|$ for $N=2$, $4$, $6$, and $8$, calculated using the weak drive equations. Parameters used here are the same as in Fig.~\ref{fig:wdl} for each $N$. Compared to $g_s^{(2)}(0)$, which is featured in the other plots, this quantity omits the effects of variations in the single-photon amplitudes. }
  \label{fig:wdl_amp_cmap}
\end{figure}

It can be proven that the sum is zero for $2p \ge N$, whereas for $2p = N-2$,
\begin{equation}
  \sum_{h=0}^{N-2} (-1)^h \binom{N-2}{h} G(N-2,N-h-1)\, G(N-2,h+1)
  = (-1)^{\frac{N}{2}+1}\frac{(N-3){!}{!}}{(N-2){!}{!}}.
  \label{SMeq:Gsum}
\end{equation}
Plugging this back into Eqs.~\eqref{SMeq:psi2_1_second_nast} and \eqref{eq:Amnsum_intermediate} yields, to lowest order in $z$ and $t$,
\begin{equation}
  \langle N-1,N-1 | \psi_1^{(2)} \rangle
  \approx \frac{(-1)^{\frac{N}{2}+1}F_1^2}{\sqrt{2}}
  \frac{(N-3){!}{!}}{(N-2){!}{!}}
  \frac{t^{N-2}}{(2z)^{N-1}}.
  \label{SMeq:psi2_1_second_last_approx}
\end{equation}
Hence, by combining Eqs.~\eqref{SMeq:psi2}, \eqref{SMeq:z2}, and \eqref{SMeq:psi2_1_second_last_approx}, we obtain an expression for the two-photon state $|\psi^{(2)}\rangle$.  The amplitude on the signal site $N-1$ is plotted in Fig.~\ref{fig:wdl_amp_cmap} for different choices of $N$.

Finally, we see that $|\psi^{(2)}\rangle = 0$ when
\begin{equation}
  \alpha \approx \frac{(-1)^{\frac{N}{2}}}{4}\frac{(N-2){!}{!}}{(N-3)!!}(2z)^{N+1},
\end{equation}
which is the UPB condition given in the main text.

The photon occupation number on the signal site $N-1$ is
\begin{align}
  n_{N-1}&=\left|\braket{N-1}{\psi^{(1)}}\right|^2\\
  &=\left|\braket{N-1}{\psi^{(1)}}\braket{N-1}{\psi^{(1)}}\right|\\
  &=\left|\sqrt{2}\braket{N-1,N-1}{\psi^{(2)}_0}\right|\\
  &=F_1^2t^{N-2}|z|^2.
\end{align}

\section{WFMC Simulations}

The Wave function Monte Carlo (WFMC) or ``quantum trajectory'' calculations \cite{BreuerBookOpen,*Carmichael_BOOK_2} reported in this work were performed using the \texttt{QuantumOptics.jl} toolbox \cite{kramer2018quantumoptics}.  The WFMC calculations take place in a tensor product space of the truncated Fock spaces for the different lattice sites.  Due to the combination of weak driving and uneven photon occupations, we found it necessary to take the following steps to ensure that the results are accurate:

\begin{itemize}
    \item Direct inclusion of the recycling terms in terms of jumps would make their rate very small, due to the low photon occupations, yet their effect may not necessarily be negligible.  Given the finite statistics generated by the Monte Carlo simulations, this would likely result in a bad sampling of rare events.  To overcome this, we note that a master equation described by a Hamiltonian $\Hami{}$ and jump operators  $=\{L_j^1,L_j^2\}=\{\sqrt{\gamma}\a_j, \sqrt{\kappa} \ad_j\a_j\}$ can be rewritten with a simultaneous transformation
    \begin{align}
        \Hami{}&\rightarrow\Hami{}'= \Hami{}+\frac{\beta}{2i}\sum_j\left(\sqrt{\gamma}\a_j+\sqrt{\kappa}\ad_j\a_j-\text{h.c.}\right) \\
        L^1_j&\rightarrow {L^1}'_j=\sqrt{\gamma}\a_j+\beta \mathds{1} \\
        L^2_j&\rightarrow {L^2}'_j=\sqrt{\kappa}\ad_j\a_j+\beta \mathds{1},
    \end{align}
where $\beta$ is an arbitrary constant. Even though this master equation generates the same evolution, the behaviour of individual trajectories can be very different as the jump rates increase from $\ev{L^\dagger L}$ to $\ev{L'^\dagger L'}$. This procedure is known as ``choosing a different unraveling'' \cite{BreuerBookOpen,*Carmichael_BOOK_2}. The limit $\beta\rightarrow\infty$ leads to a diffusion process corresponding to homodyne detecton. In our WFMC simulations, we take $\beta=0.1$.

  \item The numerical solver was run with strict tolerance settings.  For the simulations with $F=10^{-4}$, we used $\texttt{RelTol} = \texttt{AbsTol} = 10^{-18}$.  For other values of $F$, we chose similar tolerances, based on what is required for convergence.

  \item To account for slow dynamics, the state was allowed to relax freely from the vacuum to the steady state during an initial time $T_\text{relax}$, and the results were recorded over a subsequent time $T_\text{record}$.  The trajectory simulations were repeated $N_\text{traj}$ times to allow for additional averaging and estimation of statistical error through jackknife resampling. In most of the simulations, we used $T_\text{relax}=10^3\hbar$, $T_\text{record}=10^4\hbar$ and $N_\text{traj}$=10.

  \item We carefully chose the Fock cutoffs $N_\text{Fock}^{(j)}$ (i.e., the maximal amount of photons allowed on each site $j$).  Appropriate choices for each $j$ were needed to maintain accuracy while keeping the computational cost manageable. For the simulations at $F=10^{-4}$, for example, we used $N_\text{Fock}^{(j)}=3$ for all $j$; whereas to study the large-$F$ limit with $N=4$, we took $N_\text{Fock}^{(1)}=N_\text{Fock}^{(2)}=32$ and $N_\text{Fock}^{(3)}=N_\text{Fock}^{(4)}=8$.
\end{itemize}

\begin{figure}
  \includegraphics[width=0.7\linewidth]{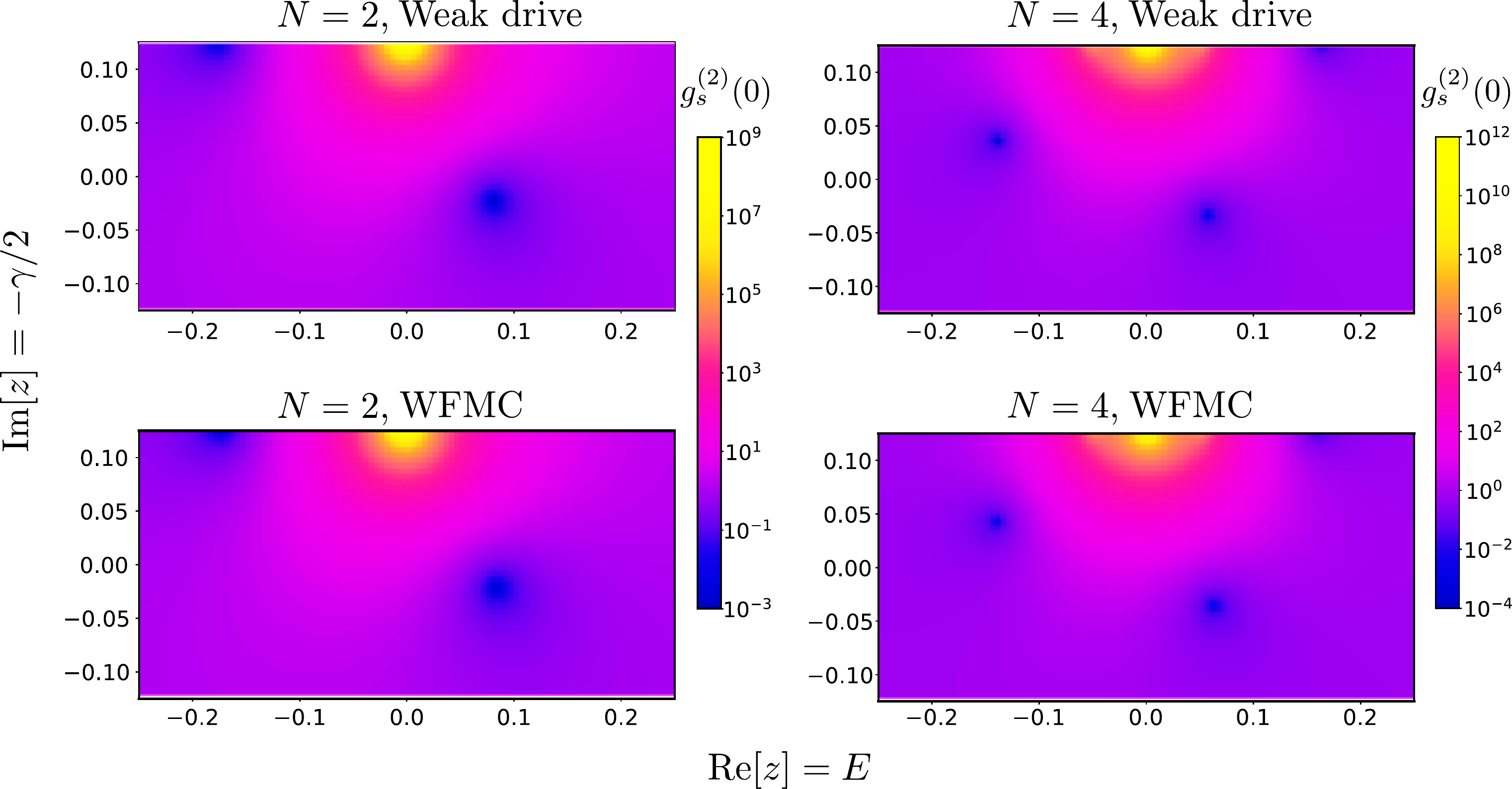}
  \caption{WFMC verification of weak drive limit result for lattice sizes $N=2$ and $N=4$.  The color of the heat map corresponds to $g_s^{(2)}(0)$.  The Kerr coefficients are $\alpha=10^{-2}$ for $N = 2$ and $2\times10^{-3}$ for $N = 4$; all other parameters are the same as in Fig.~3(a) of the main text.}
  \label{SMfig:wfmc}
\end{figure}

In all cases, we verified that the values were sufficient for proper
convergence.

\begin{figure}
  \includegraphics[width=0.45\linewidth]{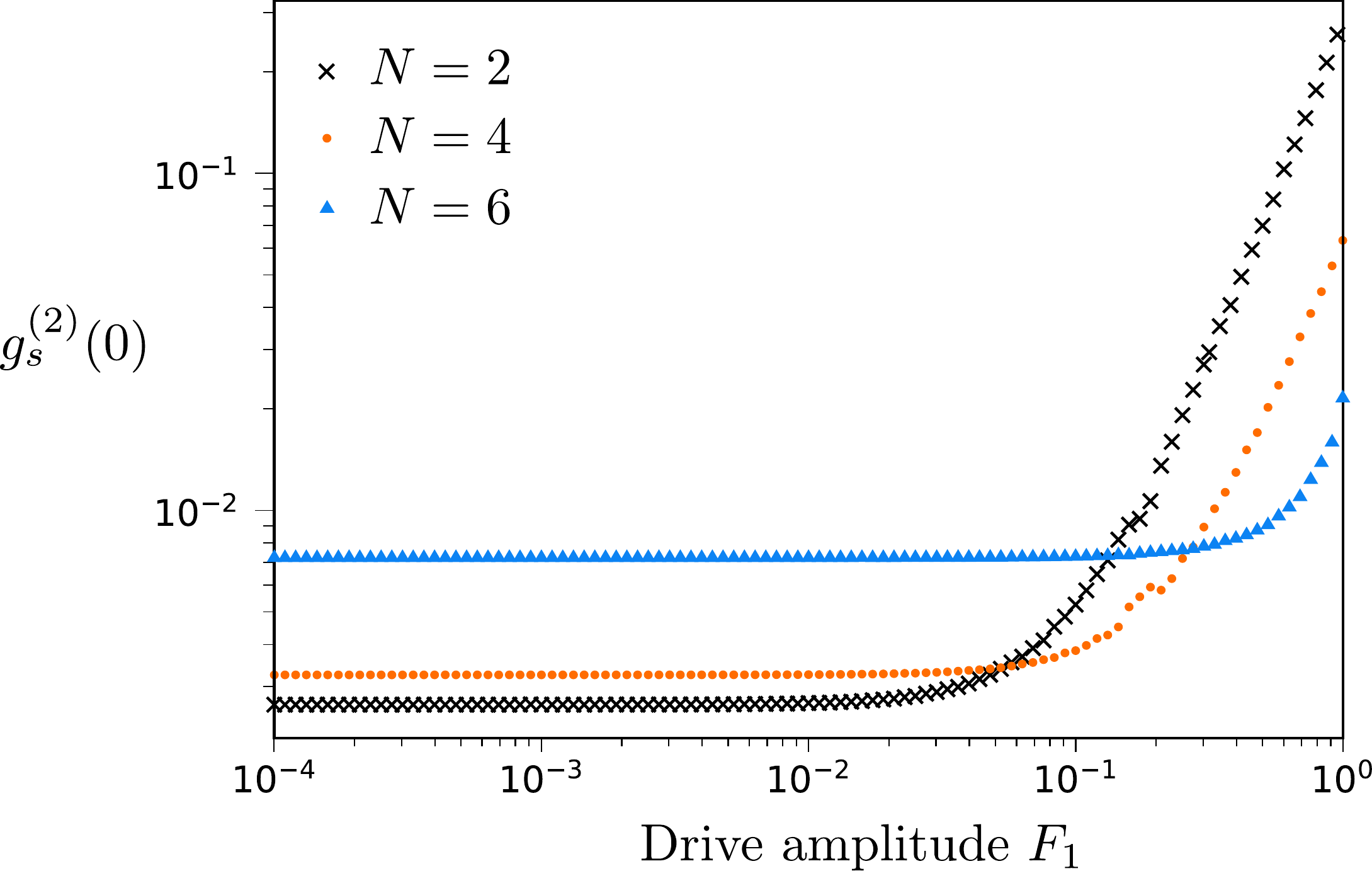}
    \caption{WFMC calculations of $g^{(2)}_s$ for different values of the drive amplitude $F_1$, with lattice sizes $N=2$, $4$, and $6$.  The other parameters are the same as in previous simulations: $t=0.1$ and $\alpha=10^{-2},2\times10^{-3},2\times10^{-4}$ for $N=2,4,6$.  The values of $z$ are chosen to be close to the optimal values for achieving UPB, as determined by searching for the minimum of $g^{(2)}_s$ on a 2D grid (due to the discretization of the search grid, for each $N$ the system is slightly detuned from the true optimum).  Pure dephasing is not included.}
    \label{fig:F_dependent}
\end{figure}

In the main text, Fig.~3(a) plots the second-order correlation $g_s^{(2)}$ shows for lattice size $N = 6$, comparing the results obtained from the weak drive equations and WFMC simulations (without pure dephasing).  In Fig.~\ref{SMfig:wfmc}, we show the analogous plots for $N = 2$ and $N = 4$.  It can be seen that the weak drive equations and WFMC results are in good agreement for different values of $N$.

In all of the other WFMC results presented in this work, we chose a drive amplitude $F_1 = 10^{-4}$ (in units of the lattice's intra-cell coupling).  The weak drive equations rely on the assumption that the drive amplitude is small.  In Fig.~\ref{fig:F_dependent}, we show the values of $g^{(2)}_s$ produced by WFMC simulations with different values of $F_1$.  It can be seen that for $F_1 \lesssim 10^{-2}$, the WFMC results become approximately independent of $F$, consistent with the existence of a well-defined weak drive limit.

\subsection{WFMC calculations of unequal-time correlation functions}

Assuming that the system reaches a steady state with constant $\ev{n_j(t)}$, the time-dependent second-order correlation function at the $j$th lattice site can be defined as
\begin{equation}\label{eq:tdepg2}
 g^{(2)}_j(\tau)=\frac{\ev{\ad_{j}(0) \ad_{j}(\tau) \a_{j}(\tau) \a_{j}(0) }}{\ev{n_j(0)}^2}.
\end{equation}

To study this numerically, we again run the simulation over a time span $T_\text{relax}$ to allow the state to relax to steady state.  Subsequently, the expectation value for the denominator of Eq.~\eqref{eq:tdepg2} is obtained over a span $T_\text{record}^\text{stat}$.  After this, the state is $\ket{\psi_s (t_0)}$, where $s=1,\ldots, N_\text{traj}$. We then define
\begin{equation}
  \ket{\phi_s(t_0)} = \frac{a_j\ket{\psi_s(t_0)}}{\norm{a_j\ket{\psi_s(t_0)}}},
\end{equation}
and propagate the system for a time $\tau$.  Then, \begin{align}\label{eq:tdepg2num}
    \ev{\ad_{j}(t_0) \ad_{j}(t_0+\tau) \a_{j}(t_0+\tau) \a_{j}(t_0) }_s&=\bra{\phi_s(t_0+\tau)}n_j\ket{\phi_s(t_0+\tau)}\norm{a_j\ket{\psi_s(t_0)}}^2\nonumber\\&=\bra{\phi_s(t_0+\tau)}n_j\ket{\phi_s(t_0+\tau)}\bra{\psi_s(t_0)}n_j\ket{\psi_s(t_0)}.
\end{align}
As we are interested in steady state properties (the time-offset $t_0$ is unimportant), the numerator of Eq.~\eqref{eq:tdepg2} is readily obtained by averaging Eq.~\eqref{eq:tdepg2num} over the different trajectories $s$.

\clearpage
\end{widetext}

\end{document}